\documentclass[twocolumn,showkeys,aps,prb,showpacs]{revtex4-1}
\usepackage{graphicx}
\usepackage[CJKbookmarks,dvipdfm,colorlinks,linkcolor=blue,citecolor=blue]{hyperref}

\begin{document}

\title{Lower lattice thermal conductivity in SbAs  than  As or Sb monolayer: a first-principles study}

\author{San-Dong Guo}
\affiliation{School of Physics, China University of Mining and
Technology, Xuzhou 221116, Jiangsu, China}

\begin{abstract}
Phonon transports of group-VA elements (As, Sb, Bi) monolayer semiconductors have been widely investigated in theory, and Sb monolayer (antimonene) of them has recently been synthesized. In this work,  phonon transport of SbAs monolayer is investigated  from a combination of  first-principles calculations and the linearized phonon Boltzmann equation. It is found that the lattice thermal conductivity of SbAs monolayer is lower than ones of both As and Sb monolayers, and  the corresponding sheet thermal conductance is 28.8 $\mathrm{W  K^{-1}}$ at room temperature. To understood lower lattice thermal conductivity in SbAs monolayer than  As and Sb monolayers, group  velocities and  phonon lifetimes of  As, SbAs and Sb monolayers are calculated. Calculated results show that  group  velocities of SbAs monolayer are between ones of As and Sb onolayers, but phonon lifetimes of SbAs are smaller than ones of both As and Sb monolayers. Hence, low lattice thermal conductivity in SbAs monolayer is attributed to very small  phonon lifetimes.
Unexpectedly, the ZA branch has very little contribution  to the total thermal conductivity, only 2.4\%,  which is obviously different from ones of As and Sb monolayers  with very large contribution.  This can be explained by very small phonon lifetimes for ZA branch of SbAs monolayer.
The large charge transfer from Sb to As atoms leads strongly polarized covalent bond, being different from As or Sb monolayer.
The strongly polarized covalent bond of SbAs monolayer  can induce  stronger  phonon anharmonicity than As or Sb monolayer,  leading to lower  lattice thermal conductivity.  We also consider  the isotope and size effects on the lattice thermal conductivity. It is
found that isotope scattering produces neglectful effect, and  the lattice thermal conductivity with the  characteristic length smaller than 30 nm can  reach a  decrease of about 47\%.
These results may offer perspectives on tuning lattice thermal conductivity by mixture of multi-elements for applications of
thermal management and thermoelectricity, and motivate further experimental efforts to synthesize monolayer SbAs.
\end{abstract}

\pacs{72.15.Jf, 71.20.-b, 71.70.Ej, 79.10.-n ~~~~~~~~~~~~~~~~~~~~~~~~~~~~~~~~~~~Email:guosd@cumt.edu.cn}

\maketitle

\section{Introduction}
Two-dimensional (2D) materials, such as  semiconducting transition-metal dichalcogenide\cite{q7}, group IV-VI\cite{q8}, group-VA\cite{q9,q10}  and group-IV\cite{q11}  monolayers, have been widely investigated both in theory and experiment, which have  potential applications in high-performance electronics, sensors and alternative energy devices.
Recently, stable semiconducting group-VA elements (As, Sb, Bi) monolayers with the  graphene-like buckled structure are  theoretically predicted  \cite{q9}, Sb monolayer (antimonene) of which  is  prepared successfully on various substrates via van der Waals epitaxy growth\cite{t8,q10}.
A stibarsen is a natural form of arsenic antimonide (SbAs), which possesses  the same layered structure with arsenic and antimony.
The counterpart monolayer material of the SbAs bulk with different typical honeycomb polymorph structures has been investigated by the first-principles calculations, and $\beta$-SbAs with a graphene-like buckled structure is predicted to be ground state by the  cohesive energies\cite{t9}.
It is found that the biaxial
tensile strain can induce semiconductor-topological insulator transition in SbAs monolayer\cite{t9}. This implies that monolayer SbAs may paly a  key role to advance the development of next generation nano-electronics, and may have wide range of potential applications
in electronic, quantum and optoelectronic devices.
\begin{figure}
  \includegraphics[width=6cm]{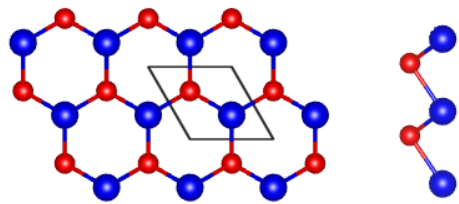}
  \caption{(Color online) Top and side views of  crystal structure
of SbAs monolayer,  and the unit cell is marked  by a black box.}\label{st}
\end{figure}

As is well known,   the  conduction of heat is  a significant
factor to determine the  performance of  nano-devices. A high thermal conductivity is beneficial to remove the accumulated heat, while  a low thermal conductivity is popular in the field of thermoelectric.  Therefore, it is interesting and necessary to investigate  thermal transport of SbAs monolayer.
In theory, thermal transports of  lots of 2D materials  have been widely investigated by semiclassical Boltzmann transport theory, Green's
function based transport techniques and equilibrium molecular dynamics simulations\cite{q21,q22,l1,l2,l4,l7,l71,l8,l9,l10,l102,l100,l101}.
The  thermal transports of group-VA elements (As, Sb, Bi) monolayers with buckled structure have been predicted\cite{l1,l2,l4},  and the  lattice thermal conductivity from As to Bi monolayer monotonously decreases. It is reported   that a buckled structure has three conflicting
effects: increasing the Debye temperature, suppressing the acoustic-optical scattering and increasing the flexural
phonon scattering.  The  former two of them can enhance lattice thermal conductivity, while the third effect can reduce one\cite{l71}.
Strain effects on lattice thermal conductivity also have been studied for various kinds of 2D materials, such as group-IV monolayers\cite{l9,l10,l100}, antimonene\cite{l102} and 2D Penta-Structures monolayers\cite{l101}. It is found that tensile strain can induce strong size effects, and the diverse strain dependence is observed, including  monotonously increasing, monotonously decreasing and  up-and-down behaviors with  strain increasing.

\begin{figure}
  \includegraphics[width=8cm]{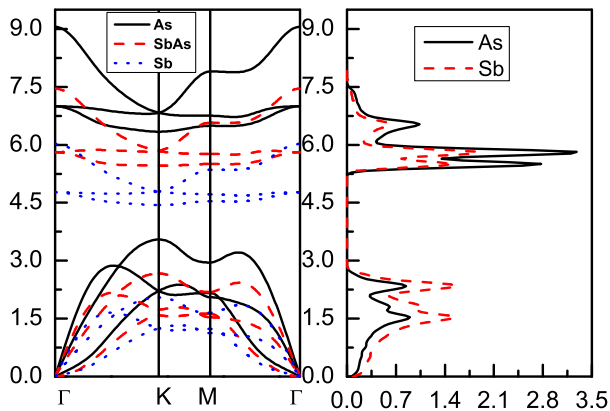}
  \caption{(Color online) Phonon band structures of As, SbAs and Sb monolayers,  together with partial DOS of SbAs monolayer.}\label{ph}
\end{figure}

In this work, the  phonon transport of SbAs monolayer is investigated based on the single-mode
RTA. It is found that  the lattice thermal conductivity of SbAs is lower than ones of both As and Sb monolayers with the same buckled honeycomb structure, which is due to shorter phonon lifetimes for SbAs than As or Sb monolayer. The shorter phonon lifetimes is due to  large charge transfer from Sb to As atoms, which can induce  strongly polarized covalent bond, leading to stronger  phonon anharmonicity than As or Sb monolayer.
 It is noted that the ZA branch of SbAs monolayer has very little contribution  to the total thermal conductivity, only 2.4\%, being obviously different from ones of As and Sb monolayers  with very large contribution. The isotope and size effects on the lattice thermal conductivity are also calculated.

The rest of the paper is organized as follows. In the next
section, we shall give our computational details about   phonon transport calculations. In the third section, we shall present  phonon transport of SbAs monolayer, together with ones of As and Sb monolayer for a comparison. Finally, we shall give our discussions and conclusions in the fourth section.
\begin{table}
\centering \caption{Lattice constants $a$  and  buckling parameter $h$  ($\mathrm{{\AA}}$);   Thermal sheet conductance $\kappa_L$ ($\mathrm{W K^{-1}}$). }\label{tab}
  \begin{tabular*}{0.48\textwidth}{@{\extracolsep{\fill}}ccccc}
  \hline\hline
Name& $a$ & $h$ &  $\kappa_L$\\\hline\hline
As&3.61 (3.61\cite{l2,l71})&1.40 (1.40\cite{l71})   &161.1 (234.8\cite{l2},127.4\cite{l71})   \\\hline
SbAs&3.87 (3.86\cite{t9})&1.52 (1.52\cite{t9})   &28.8\\\hline
Sb&4.12 (4.12\cite{l2,l71})&1.65 (1.64\cite{l71})  &46.6 (51.9\cite{l2},29.6\cite{l71})\\\hline\hline
\end{tabular*}
\end{table}

\section{Computational detail}
First-principles calculations are performed within projector augmented-wave method using the VASP code\cite{pv1,pv2,pv3,pbe}.
The generalized gradient approximation   of Perdew-Burke-Ernzerhof (PBE-GGA) is adopted as exchange-correlation functional\cite{pbe}.
 The unit cells  of As, SbAs and Sb monolayers   are  built with the vacuum region of larger than 16 $\mathrm{{\AA}}$ to avoid spurious interaction.
 A 20 $\times$ 20 $\times$ 6 q-mesh is used during structural relaxation with  a Hellman-Feynman force convergence threshold of $10^{-4}$ eV/ $\mathrm{{\AA}}$.
 A plane-wave basis set is employed with kinetic energy cutoff of 400 eV.
 The energy convergence criterion is
used, and  when the energy difference is less than $10^{-8}$ eV, the self-consistent calculations are
considered to be converged.
The lattice thermal conductivity   is performed with the single mode RTA and linearized phonon Boltzmann equation,  as implemented in the Phono3py code\cite{pv4}.
The interatomic force constants (IFCs) are calculated by
the finite displacement method.
The second order harmonic and third
order anharmonic IFCs  are
calculated by using a 5 $\times$ 5 $\times$ 1  supercell  and a  4 $\times$ 4 $\times$ 1 supercell, respectively.
Using the harmonic IFCs, phonon dispersion can be calculated by Phonopy package\cite{pv5}, determining  the allowed three-phonon scattering processes, group velocity  and specific heat.
 Based on third-order anharmonic IFCs, the phonon lifetimes can be attained from the three-phonon scattering rate. To compute lattice thermal conductivities, the reciprocal spaces of the primitive cells  are sampled using the 50 $\times$ 50 $\times$ 2 meshes.
For 2D material, the calculated  lattice  thermal conductivity  depends on the length of unit cell used in the calculations along z direction\cite{2dl}.  The lattice  thermal conductivity should be normalized by multiplying $Lz/d$, in which  $Lz$ is the length of unit cell along z direction  and $d$ is the thickness of 2D material, but the $d$  is not well defined.   In this work, the length of unit cell (18 $\mathrm{{\AA}}$) along z direction is used as the thickness of As, SbAs and Sb monolayers. To make a fair comparison between various 2D monolayers, the thermal sheet conductance can be used, defined as $\kappa$ $\times$ $d$.

\section{MAIN CALCULATED RESULTS AND ANALYSIS}
The  $\beta$-SbAs monolayer with a graphenelike buckled honeycomb structure is similar to
those of group-VA and group-IV monolayers, and the schematic crystal structure is shown in \autoref{st}. The SbAs monolayer (No.156) has
 lower symmetry compared with As and Sb monolayers (No.164) due to the different
types of atoms constituting the compound.
Firstly, the lattice constants are optimized, and the resulting $a$=$b$=3.867 $\mathrm{{\AA}}$ and buckling parameter $h$=1.515 $\mathrm{{\AA}}$, which are very close to previous  theoretical values\cite{t9}.  The lattice constants and buckling parameter $h$ in this or previous works\cite{l2,l71,t9} are summarized in \autoref{tab} for As, SbAs and Sb monolayers. It is expected that both lattice constants and buckling parameter of SbAs monolayer are greater than ones of As monolayer, but less than ones of Sb monolayer.

The phonon dispersions of As, SbAs and Sb monolayers in high symmetry directions  are plotted  in \autoref{ph}, together with partial  density of states (DOS) of SbAs monolayer. Due to  containing  two  atoms in their unit cell,  3 acoustic
and 3 optical phonon branches can be observed.
 Due to buckling of As, SbAs and Sb monolayers,  their out-of-plane acoustic modes  have coupling with the in-plane longitudinal acoustic (LA) and transversal acoustic (TA) modes. However, the out-of-plane acoustic modes are still marked with  ZA  modes. It is clearly seen that  LA and TA branches are linear near the $\Gamma$ point, while ZA branch  deviates from linearity, which shares the general feature of 2D materials\cite{q21,q22,l1,l2,l4,l7,l71,l8,l9,l10,l102,l100,l101}.
It is clearly seen that both  acoustic and  optical branches overall move downward from As to SbAs to Sb monolayer, and the  widthes of  acoustic and  optical branches gradually become narrow.  The maximal acoustic vibration frequency (MAVF)   is 3.55 THz, 2.67 THz and 2.05 THz from As to SbAs to Sb monolayer, and the width of optical branches (WOOB)  is 2.71 THz, 1.99 THz and 1.58 THz, respectively. The low MAVF and narrow WOOB  mean small group velocities, which is benefit to low thermal conductivity. The Debye temperature $\theta_D$ can be calculated with the
MAVF by  $\theta_D=h\nu_m/k_B$\cite{q21,q22}, where $h$, $k_B$ are the Planck constant and  Boltzmann
constant.  The calculated $\theta_D$ for As, SbAs and Sb monolayers is 170.36, 128.13 and 98.38 K, respectively.
A phonon band gap between acoustic and optical branches can be observed, and the corresponding value is 2.79 THz, 2.79 THz and 2.39 THz from As to SbAs to Sb monolayer.  The gaps of As and Sb monolayers are  due to the violation of  reflection symmetry selection
rule in the harmonic approximation\cite{l71}, while the gap of SbAs is due to not only the violation of  reflection symmetry selection
rule but also  mass differences between the constituent atoms\cite{m1-1,m3-1}. Therefore, the gap of As monolayer is the same with that of SbAs monolayer, not like that the gap gradually decreases from As to Sb to Bi monolayer.  According to partial DOS of SbAs monolayer, the  optical modes of  SbAs are mainly
from As vibrations, while the  acoustic branches are mainly  due to the vibrations of Sb.
\begin{figure}[!htb]
  \includegraphics[width=8cm]{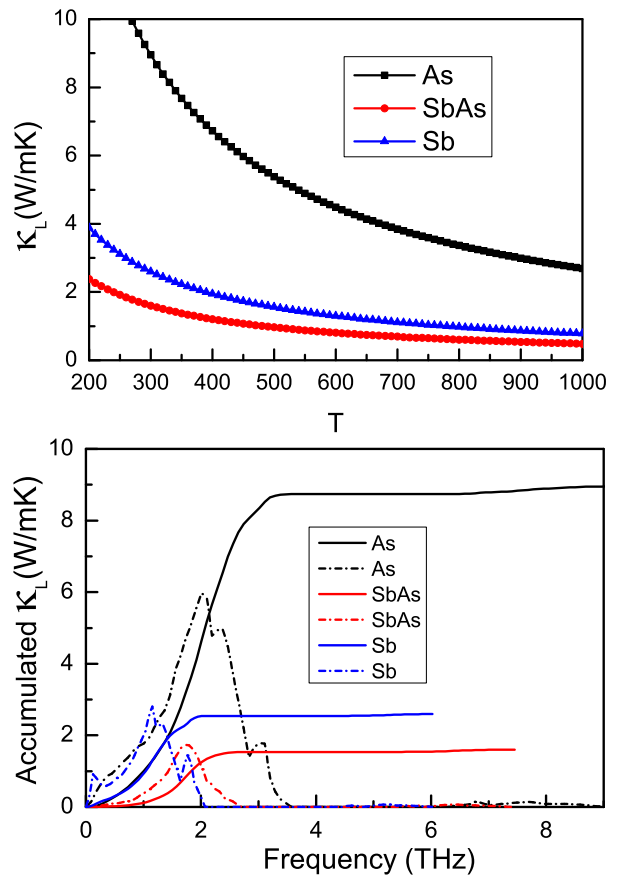}
  \caption{(Color online) Top: the lattice thermal conductivities of As, SbAs and Sb monolayers  as a function of temperature. Bottom: the accumulated lattice thermal conductivities with respect to frequency (solid lines), and the derivatives (short dash dot lines).}\label{kl}
\end{figure}
\begin{figure}[!htb]
  \includegraphics[width=8cm]{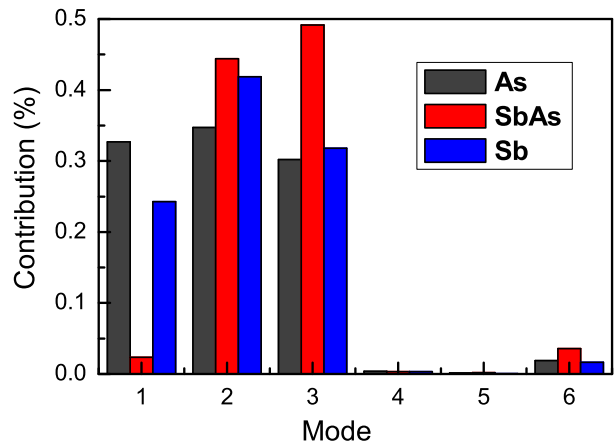}
  \caption{(Color online) The phonon modes contributions toward total lattice thermal conductivity of As, SbAs and Sb monolayers (300 K).}\label{mode}
\end{figure}
\begin{figure}
  \includegraphics[width=7cm]{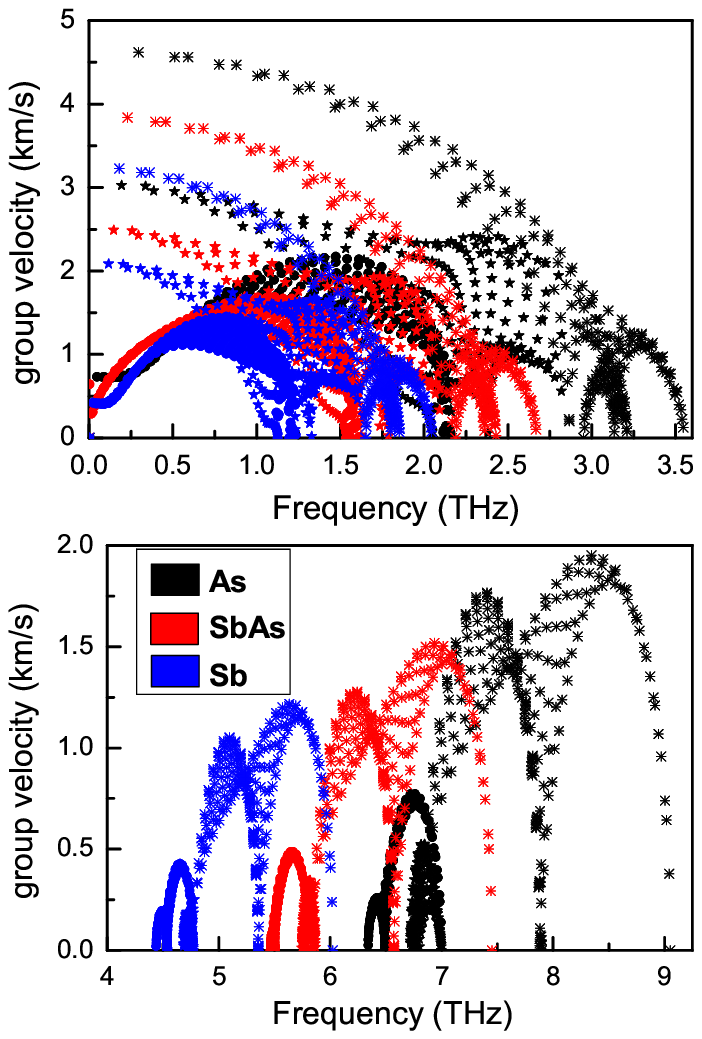}
  \caption{(Color online) The phonon mode group velocities of  As, SbAs and Sb monolayers in the first Brillouin zone. Top: the ZA (circle symbol), TA (star symbol) and LA (star ($\ast$) symbol) acoustic branches; Bottom: the first  (circle symbol), second (star symbol) and third (star ($\ast$) symbol) optical branches. }\label{v}
\end{figure}
\begin{figure}
  \includegraphics[width=7cm]{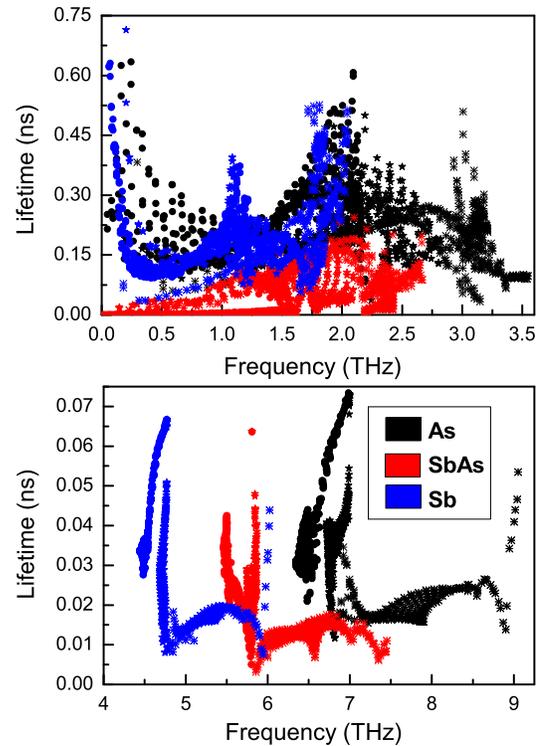}
  \caption{(Color online) The phonon mode lifetimes of As, SbAs and Sb monolayers in the first Brillouin zone. Top: the ZA (circle symbol), TA (star symbol) and LA (star ($\ast$) symbol) acoustic branches; Bottom: the first  (circle symbol), second (star symbol) and third (star ($\ast$) symbol) optical branches.}\label{t}
\end{figure}

The lattice  thermal conductivities of  As, SbAs and Sb monolayers are calculated with RTA method, which  are shown in \autoref{kl}. The room-temperature lattice  thermal conductivity of As, SbAs and Sb monolayers is 8.95 $\mathrm{W m^{-1} K^{-1}}$, 1.60 $\mathrm{W m^{-1} K^{-1}}$  and 2.59 $\mathrm{W m^{-1} K^{-1}}$, respectively, with the same thickness $d$ (18 $\mathrm{{\AA}}$). To make a fair comparison between various 2D materials, their  thermal sheet conductance\cite{2dl} is  161.1 $\mathrm{W K^{-1}}$,  28.8  $\mathrm{W K^{-1}}$  and  46.6 $\mathrm{W K^{-1}}$, respectively. The room-temperature thermal sheet conductances are listed in \autoref{tab}, together with available previous theoretical values of As and Sb monolayers\cite{l2,l71}, which have been converted into thermal sheet conductances. Our calculated values  are  in the range of previous ones.
 The SbAs monolayer can be attained by  using As (Sb) atoms to replace one sublayer of Sb (As) monolayer. It is expected that the  lattice  thermal conductivity of SbAs monolayer should be between ones of As and Sb monolayers. However, it is clearly seen that the  lattice  thermal conductivity of SbAs monolayer is lower than any of ones of As and Sb monolayers.
The cumulative lattice thermal conductivity  and the derivatives  are also plotted in \autoref{kl} at room temperature.
The acoustic branches of As, SbAs and Sb monolayers provide a contribution of 97.6\%, 95.9\% and 98.0\%, respectively, which meets the usual picture that acoustic branches  dominate lattice thermal conductivity. It is found that the slope of cumulative lattice thermal conductivity of SbAs monolayer is smaller than ones of As and Sb monolayers, which means that the low-frequency acoustic  phonons of SbAs monolayer  have very little contribution to thermal conductivity.
To further examine the relative contributions of every phonon  modes to the total lattice thermal conductivity, the phonon modes contributions toward total lattice thermal conductivity of As, SbAs and Sb monolayers (300 K) are shown in \autoref{mode}.
It is found that ZA branch  provides very large  contribution for As and Sb monolayers, but  very little contribution for SbAs monolayer, only 2.4\%.
However, the TA and LA branches have larger   contribution for SbAs monolayer than  As and Sb monolayers.  Especially, the LA branch provides almost half of lattice  thermal conductivity  for SbAs monolayer, up to 49.1\%. It is clearly seen that the third optical branch has relatively obvious contribution.

To identify the underlying mechanism of lower lattice thermal conductivity in SbAs monolayer than  As and Sb monolayers,
 phonon mode group velocities  of  As, SbAs and Sb monolayers are calculated,  which are shown in \autoref{v}.
For all branches, it is clearly seen that  most of group velocities  become small from As to SbAs to Sb monolayer, which can lead to decreasing  lattice thermal conductivity. From As to SbAs to Sb monolayer, the largest  group velocity   changes from  2.17  $\mathrm{km s^{-1}}$ to 1.69  $\mathrm{km s^{-1}}$ to 1.45  $\mathrm{km s^{-1}}$ for ZA branch, from 3.03  $\mathrm{km s^{-1}}$ to 2.49 $\mathrm{km s^{-1}}$ to 2.08 $\mathrm{km s^{-1}}$ for TA branch , from 4.62 $\mathrm{km s^{-1}}$ to 3.84 $\mathrm{km s^{-1}}$ to 3.23 $\mathrm{km s^{-1}}$ for LA branch. Therefore, the changes of  group velocities can not explain lower lattice thermal conductivity in SbAs monolayer than  As and Sb monolayers.
In the single-mode RTA method,  phonon  lifetimes are merely proportional to lattice thermal conductivity \cite{pv4},
which can be calculated by  three-phonon scattering rate from  third-order anharmonic IFCs.
The  phonon  lifetimes of  As, SbAs and Sb monolayers are plotted  in \autoref{t}.
For all phonon branches, most of phonon  lifetimes  become large from SbAs to Sb to As monolayer, which means that SbAs monolayer has lowest  lattice thermal conductivity. It is clearly seen that  most of phonon  lifetimes of ZA branch for SbAs monolayer are very small, which leads to very little contributions toward total lattice thermal conductivity of SbAs monolayer for ZA branch.
Because acoustic branches have larger group velocities and phonon  lifetimes than optical ones,  acoustic branches dominate the lattice thermal conductivity.
According to group velocities,  the lattice thermal conductivity of SbAs monolayer should be between ones of As and Sb monolayers. However, SbAs monolayer has the lowest one by analyzing phonon  lifetimes.
So, the short phonon lifetimes  lead to  lower lattice thermal conductivity in SbAs monolayer than As and Sb monolayers.
\begin{figure}
  \includegraphics[width=7.0cm]{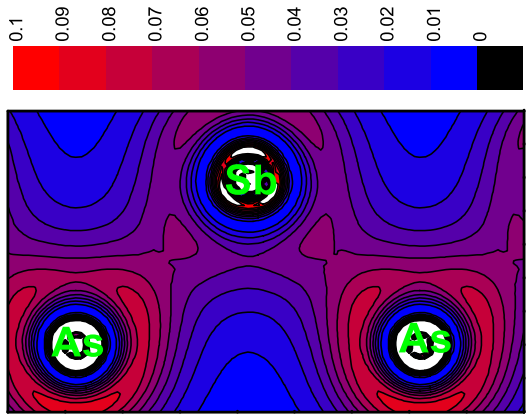}
  \caption{(Color online) The charge density distributions  of SbAs (unit:$\mathrm{|e|}$/$\mathrm{bohr^3}$). }\label{den}
\end{figure}

\begin{figure}
  \includegraphics[width=8.0cm]{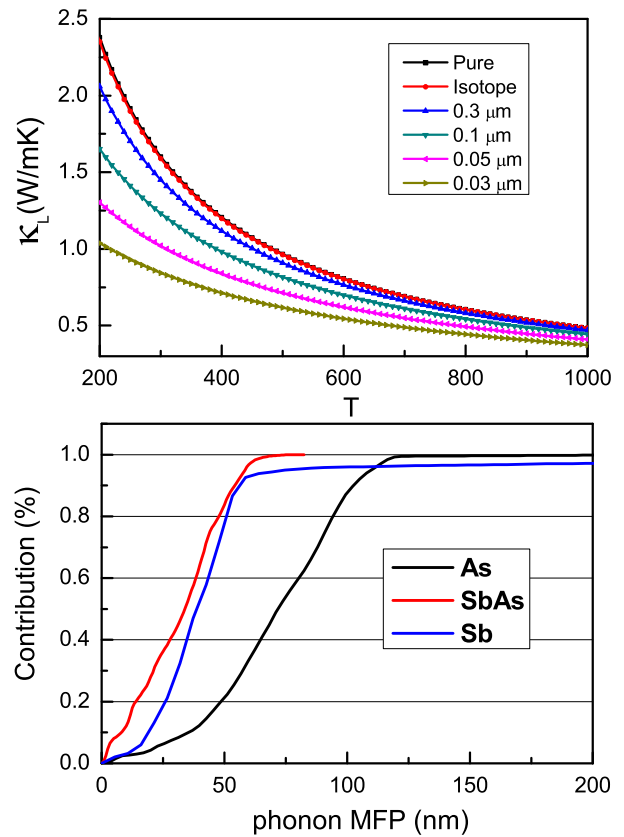}
  \caption{(Color online)Top: the lattice thermal conductivities  of infinite (Pure and Isotope) and finite-size (0.3, 0.1, 0.05 and 0.03 $\mathrm{\mu m}$) monolayer SbAs as a function of temperature; Bottom: the cumulative lattice thermal conductivity of As, SbAs and Sb monolayers  divided by their total lattice thermal conductivity with respect to phonon MFP at room temperature.}\label{mkl}
\end{figure}

The distribution of electrons in real space can be described by the charge density distributions, which determines the bond characteristics between atoms.
The charge density distributions  of SbAs monolayer is shown in \autoref{den}.  For As
and Sb monolayers, there is no charge transfer among As or Sb atoms due to  the same atom types to form bond.
But for SbAs monolayer, from Sb  to As atom, the charge density distinctly increases, which indicates that charge transfer between Sb and As atoms occurs.
This  can be understood by
considering the different electronegativity between As (2.18) and Sb (2.05)
atoms.  The charge transfer from Sb to As atom can give rise to the relatively strongly polarized
covalent bond. The inhomogeneous distribution of
charge density of monolayer SbAs can induce larger  anharmonicity  than As and Sb monolayers,  which leads to stronger intrinsic phonon-phonon scattering, driving the lower lattice thermal conductivity of monolayer SbAs than As and Sb monolayers. Similar results can also be found in SiC and GaN monolayers\cite{q22,gsd}

According to the formula given  by Shin-ichiro Tamura\cite{q24}, the phonon-isotope scattering  is considered. Calculated results show very insensitivity of lattice thermal conductivity to isotopes, as shown in \autoref{mkl}.
A most simple boundary scattering treatment is adopted, whose
 scattering rate can be attained by $v_g/L$, where $v_g$, $L$  are the group velocity and  the boundary MFP, respectively.
The lattice thermal conductivities  of finite-size (0.3, 0.1, 0.05 and 0.03 $\mathrm{\mu m}$) SbAs monolayer as a function of temperature are plotted in \autoref{mkl}. As the sample size of SbAs decreases ,  the lattice thermal conductivity decreases, which is  due to enhanced boundary scattering.
For the 0.3, 0.1, 0.05 and 0.03 $\mathrm{\mu m}$ cases,  the room-temperature lattice thermal conductivity of SbAs monolayer  is reduced by about 9.50\%, 23.13\%,  36.33\% and 47.40\%  compared with infinite (Pure) case.

The thermal conductivity spectroscopy technique  can measure MFP distributions over
a wide range of length scales\cite{pl1}. Therefore, cumulative lattice thermal conductivity divided by total lattice thermal conductivity  of As, SbAs and Sb monolayers with respect to phonon MFP, at room temperature,  are plotted in in \autoref{mkl}, which can be used to study size effects in heat conduction.
As the  MFP increases, the cumulative lattice thermal conductivity divided by total lattice thermal conductivity increases, and then approaches one.
From As to SbAs to Sb monolayer, the corresponding critical  MFP    changes from 124 nm  to 68 nm to 432 nm. It has been proved that  strain can induce very  strong size effects on lattice thermal conductivity  of Sb monolayer\cite{l102}, which means that  critical  MFP   significantly depends on lattice constants.
When the lattice thermal conductivity is  reduced to 60\% by nanostructures,  the characteristic length varies  from 80 nm to 38 nm to 44 nm  from As to SbAs to Sb monolayer.

\section{Discussions and Conclusion}
 Compared with the planar geometry of graphene, it is demonstrated that a buckled structure has conflicting effects on  lattice thermal conductivity\cite{l71}. On the one hand, buckled structure can increase lattice thermal conductivity by the formation of an acoustic-optical gap,
suppressing A+A$\longleftrightarrow$O scattering.  On the other hand, it can reduce one by  breaking  the out-of-plane symmetry, increasing anharmonic phonon scattering. The As, SbAs and Sb monolayers all possess  buckled structure.   The buckling parameter $h$  monotonically  increases from As to SbAs to Sb monolayer, and the acoustic-optical gap of SbAs monolayer is larger than that of Sb monolayer.
Therefore, the trend of lattice thermal conductivity from As to SbAs to Sb monolayer  cannot be addressed by conflicting effects caused by buckled structure.
Based on  the formula proposed  by Slack\cite{m1},    the  average atomic mass,  interatomic bonding, crystal structure and  anharmonicity
determine the lattice thermal conductivity of a material.  Calculated results show that the anharmonicity can explain the trend of  lattice thermal conductivity from As to SbAs to Sb monolayer. According to \autoref{den}, the charge transfer of monolayer SbAs can induce larger  anharmonicity  than As and Sb monolayers,  which produces  stronger intrinsic phonon-phonon scattering, leading to shorter phonon lifetimes (\autoref{t}). This drives  the lower lattice thermal conductivity of monolayer SbAs than As and Sb monolayers.

Strain  has been proved to be very effective to tune lattice thermal conductivities of various 2D materials\cite{l9,l10,l100,l102,l101}. For planar structure, for example  graphene, the reflection symmetry selection rule strongly restricts anharmonic phonon scattering, leading to very high lattice thermal conductivity \cite{l71}. Tensile strain can make a buckled structure turn into a  planar structure like penta-$\mathrm{SiN_2}$, predicted by the first-principles calculations\cite{l101}. When the  structure becomes perfectly planar, the lattice thermal conductivity of  penta-$\mathrm{SiN_2}$ suddenly jumps up by 1 order of magnitude. In fact, the  buckled  structure of Bi monolayer  becomes perfectly planar, which has been  experimentally achieved\cite{m2}.
Therefore, it is possible for SbAs monolayer to achieve  high thermal conductivity by tensile strain,  which can
dissipate heat efficiently in SbAs monolayer-based nano-electronics devices. The As, Sb, Bi and SbAs monolayers with the  graphene-like buckled structure are predicted to stable  in theory\cite{q9,t9}.  The Sb monolayer (antimonene) has been achieved in experiment\cite{t8,q10}, and it has been proved that tensile strain can enhance structural stability\cite{l102} for Sb monolayer, which  provides  guidance on
fabrication of these monolayers.  The Bi monolayer has been  successfully  synthesized by tesile strain\cite{m2}. Therefore, it is possible to achieve SbAs monolayer in experiment by tensile strain.

In summary,  the lattice thermal conductivities of SbAs monolayer, together with As and Sb monolayers,
 are  performed by solving  the linearized phonon Boltzmann equation within
the single-mode RTA. Calculated results show that the lattice thermal conductivity of SbAs is lower than ones of both As and Sb monolayers, which is due to the shorter phonon lifetimes for SbAs than As or Sb monolayer. The short phonon lifetimes is due to charge transfer from Sb to As atoms, being different from  that in As or Sb monolayer, which  leads to  strong intrinsic phonon-phonon scattering.
The size dependence of lattice thermal conductivity  is studied, which can  provide  guidance on
designing nanostructures.
 This work provides insight into  phonon transport in SbAs monolayer, and  offers new idea on tuning lattice thermal conductivity by mixture of multi-elements.

\begin{acknowledgments}
This work is supported by the National Natural Science Foundation of China (Grant No.11404391). We are grateful to the Advanced Analysis and Computation Center of CUMT for the award of CPU hours to accomplish this work.
\end{acknowledgments}

\end{document}